\PassOptionsToPackage{unicode}{hyperref}
\PassOptionsToPackage{hyphens}{url}
\PassOptionsToPackage{dvipsnames,svgnames,x11names}{xcolor}
\documentclass[12pt]{article}

\usepackage{amsmath,amssymb}
\usepackage{iftex}
\ifPDFTeX
  \usepackage[T1]{fontenc}
  \usepackage[utf8]{inputenc}
  \usepackage{textcomp} 
\else 
  \usepackage{unicode-math}
  \defaultfontfeatures{Scale=MatchLowercase}
  \defaultfontfeatures[\rmfamily]{Ligatures=TeX,Scale=1}
\fi
\usepackage{lmodern}
\ifPDFTeX\else  
\fi
\IfFileExists{upquote.sty}{\usepackage{upquote}}{}
\IfFileExists{microtype.sty}{
  \usepackage[]{microtype}
  \UseMicrotypeSet[protrusion]{basicmath} 
}{}
\makeatletter
\@ifundefined{KOMAClassName}{
  \IfFileExists{parskip.sty}{%
    \usepackage{parskip}
  }{
    \setlength{\parindent}{0pt}
    \setlength{\parskip}{6pt plus 2pt minus 1pt}}
}{
  \KOMAoptions{parskip=half}}
\makeatother
\usepackage{xcolor}
\makeatletter
\ifx\paragraph\undefined\else
  \let\oldparagraph\paragraph
  \renewcommand{\paragraph}{
    \@ifstar
      \xxxParagraphStar
      \xxxParagraphNoStar
  }
  \newcommand{\xxxParagraphStar}[1]{\oldparagraph*{#1}\mbox{}}
  \newcommand{\xxxParagraphNoStar}[1]{\oldparagraph{#1}\mbox{}}
\fi
\ifx\subparagraph\undefined\else
  \let\oldsubparagraph\subparagraph
  \renewcommand{\subparagraph}{
    \@ifstar
      \xxxSubParagraphStar
      \xxxSubParagraphNoStar
  }
  \newcommand{\xxxSubParagraphStar}[1]{\oldsubparagraph*{#1}\mbox{}}
  \newcommand{\xxxSubParagraphNoStar}[1]{\oldsubparagraph{#1}\mbox{}}
\fi
\makeatother

\usepackage{longtable,booktabs,array}
\usepackage{calc} 
\usepackage{etoolbox}
\makeatletter
\patchcmd\longtable{\par}{\if@noskipsec\mbox{}\fi\par}{}{}
\makeatother
\IfFileExists{footnotehyper.sty}{\usepackage{footnotehyper}}{\usepackage{footnote}}
\makesavenoteenv{longtable}
\usepackage{graphicx}
\makeatletter
\def\maxwidth{\ifdim\Gin@nat@width>\linewidth\linewidth\else\Gin@nat@width\fi}
\def\maxheight{\ifdim\Gin@nat@height>\textheight\textheight\else\Gin@nat@height\fi}
\makeatother
\setkeys{Gin}{width=\maxwidth,height=\maxheight,keepaspectratio}
\makeatletter
\def\fps@figure{htbp}
\makeatother

\makeatletter
\@ifpackageloaded{caption}{}{\usepackage{caption}}
\AtBeginDocument{%
\ifdefined\contentsname
  \renewcommand*\contentsname{Table of contents}
\else
  \newcommand\contentsname{Table of contents}
\fi
\ifdefined\listfigurename
  \renewcommand*\listfigurename{List of Figures}
\else
  \newcommand\listfigurename{List of Figures}
\fi
\ifdefined\listtablename
  \renewcommand*\listtablename{List of Tables}
\else
  \newcommand\listtablename{List of Tables}
\fi
\ifdefined\figurename
  \renewcommand*\figurename{Figure}
\else
  \newcommand\figurename{Figure}
\fi
\ifdefined\tablename
  \renewcommand*\tablename{Table}
\else
  \newcommand\tablename{Table}
\fi
}
\@ifpackageloaded{float}{}{\usepackage{float}}
\floatstyle{ruled}
\@ifundefined{c@chapter}{\newfloat{codelisting}{h}{lop}}{\newfloat{codelisting}{h}{lop}[chapter]}
\floatname{codelisting}{Listing}

\makeatother
\makeatletter
\@ifpackageloaded{caption}{}{\usepackage{caption}}
\@ifpackageloaded{subcaption}{}{\usepackage{subcaption}}
\makeatother

\ifLuaTeX
  \usepackage{selnolig}  
\fi
\usepackage[]{natbib}
\usepackage{bookmark}
\usepackage{bm}
\usepackage{tabularx}
\usepackage{multirow}
\usepackage{hyperref}

\IfFileExists{xurl.sty}{\usepackage{xurl}}{} 
\hypersetup{
  pdftitle={Title},
  pdfauthor={Author 1; Author 2},
  pdfkeywords={3 to 6 keywords, that do not appear in the title},
  colorlinks=true,
  linkcolor={blue},
  filecolor={Maroon},
  citecolor={Blue},
  urlcolor={Blue},
  pdfcreator={LaTeX via pandoc}}

\newcommand{\anon}{1}

\begin{document}

\def\spacingset#1{\renewcommand{\baselinestretch}%
{#1}\small\normalsize} \spacingset{1}


\if1\anon
{
  \title{\bf Bayesian Classification with Probit-link Split-and-merge Gaussian Process Prior in EEG-based Brain-Computer Interfaces}
  \author{Yunong Wu\\
    Department of Statistics, Indiana University,\\
    Jane E. Huggins \\
    Department of Physical Medicine and \\Rehabilitation and Department of Biomedical Engineering, \\University of Michigan\\
    Jian Kang\\
    Department of Biostatistics, University of Michigan\\
    Tianwen Ma\thanks{
    The authors gratefully acknowledge ...}
    \hspace{.2cm}\\
    Department of Biostatistics and Bioinformatics, Emory University\\
    }
  \maketitle
} \fi

\if0\anon
{
  \bigskip
  \bigskip
  \bigskip
  \begin{center}
    {\LARGE\bf Bayesian Classification with Probit-link Split-and-merge Gaussian Process Prior in EEG-based Brain-Computer Interfaces}
\end{center}
  \medskip
} \fi

\bigskip
\begin{abstract}
A Brain-Computer Interface (BCI) speller systems based on Event-Related Potentials (ERPs) enables users to select characters by detecting brain responses to visual stimuli, recorded through electroencephalogram (EEG). One challenge is to accurately identify target-related responses, such as the P300 component. However, existing methods tend to ignore feature selection, perform feature selection without interpretability, or require large computational effort or data manipulation. To address these limitations, we propose a novel Bayesian generative modeling framework to the binary classification of EEG responses to stimuli. Our approach employs a Probit-link Split-and-merge Gaussian Process (P-SMGP) prior to perform spatial-temporal feature selection, effectively capturing the distinctions between target and non-target ERP responses. Through both simulation studies and real EEG data analysis, our approach provides statistical interpretations on 
transformed ERP functions
while maintaining comparable prediction accuracy with a computationally motivated design. These findings underscore the value of interpretable, stimulus-level modeling for advancing predictive and personalized BCI systems.
\end{abstract}

\noindent%
{\it Keywords:} Brain-Computer Interface, BCI, Bayesian methods, Gaussian Process, Feature Selection
\vfill

\newpage
\spacingset{1.8} 

\section{Introduction}
\label{sec-intro}

\begin{figure}[ht] 
  \centering 
  \includegraphics[width=0.8\textwidth]{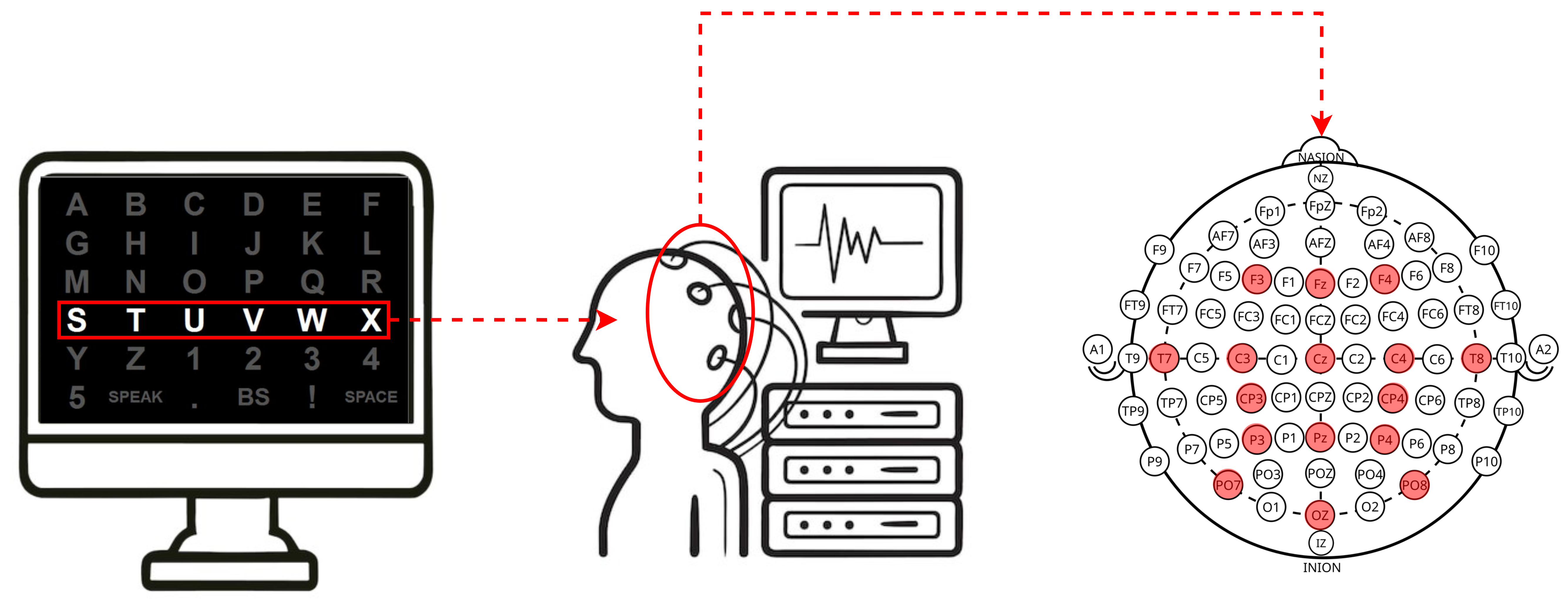} 
  \caption{\small A simple illustration of the P300 ERP-based BCI design. The participant wearing an EEG cap is asked to face a virtual keyboard of $6\times6$ grid in the left column. While the virtual keyboard is randomly highlighting rows and columns, the human brain responds to the external stimuli, and the signals are recorded by a EEG device. A computer analyzes and interprets the EEG recordings and send the feedback to the virtual screen. The right column is a figure illustrating a 64-channel electrode placement based on the International 10-20 system \cite{jasper_ten-twenty_1958}.}
  \label{fig:1} 
\end{figure}

The Brain-Computer Interface (BCI) offers a direct communication pathway between the brain and external devices, enabling users to control technology via neural activity. The EEG-based BCI device is widely used due to their non-invasiveness, cost-effectiveness, and high temporal resolution \cite{niedermeyer_electroencephalography_2005}.
One of the most popular applications of EEG-based BCIs is the P300 Event-Related Potential (ERP)-based speller system \cite{farwell_talking_1988}. This technology detects ERPs embedded in the electroencephalogram (EEG) and helps the people with severe neuromuscular diseases, such as amyotrophic lateral sclerosis (ALS), communicate. An ERP is defined as a change of brain activity in response
to an external event. The P300 component is characterized by a positive voltage deflection approximately 300 ms after the presentation of an infrequent but relevant stimulus \cite{rodden_brief_2008}. The conventional P300 speller system employs the row-and-column paradigm (RCP) developed by \citet{farwell_talking_1988}. 
In the 6×6 grid design, a stimulus refers to a single visual event in which one entire row or one entire column is flashed on the screen. A sequence refers to a full cycle of stimuli in which all six rows and all six columns are each flashed exactly once, typically in a randomized order—i.e., 12 stimuli per sequence. For a given character, exactly two of the 12 stimuli in each sequence (the flashed row and the flashed column that contain the desired character) are supposed to elicit the P300 ERP responses, while the remaining are not.
The appearance of a target stimulus elicits a P300 ERP response from the user while the appearance of a non-target stimulus does not. The system computes the classifier score for each stimulus, identify the row and column with the maximum scores, and picks the intended character as the intersection of the one row and one column. In practice, this process is repeated with multiple replications, and stimulus-specific scores are summed up across replications for the final character selection. Figure \ref{fig:1} illustrates the process in brief.

\begin{figure}[ht] 
  \centering 
  \includegraphics[width=0.6\textwidth]{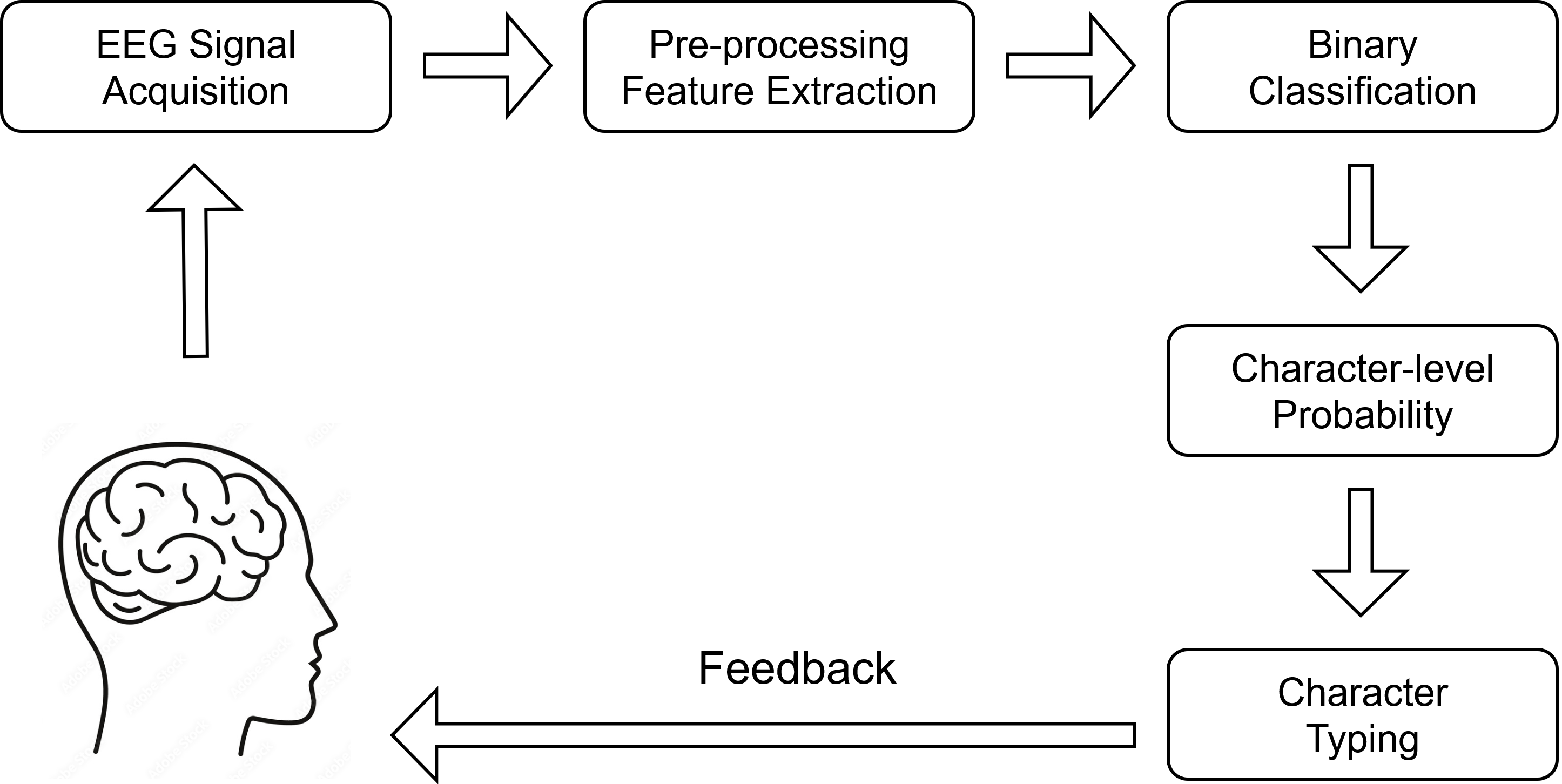} 
  \caption{\small The conventional framework of P300 ERP-based BCI speller. The process starts with data collection via neuro-physiological sensors to record raw EEG signals. Signal pre-processing and feature extraction are applied to raw signals. A binary classification is performed to compute the stimulus-specific classifier scores and the character-level probability. Finally, the intended key is selected by identifying the most plausible row and column and displayed to the user.}
  \label{fig:2} 
\end{figure}

A conventional P300 ERP-based BCI speller follows a standard closed-loop pipeline—from EEG acquisition and pre-processing, to feature extraction, binary target–non-target classification, and character decision via aggregating row/column evidence across repeated sequences (Figure~\ref{fig:2}).

Feature selection is necessary to ensure reliable target--non-target discrimination in the presence of signal overlap and low signal-to-noise ratios. Existing approaches can be broadly grouped into three categories. First, many commonly used discriminative classifiers---including logistic regression (LR) \cite{viana_logistic_2014}, support vector machines (SVM) \cite{kaper_bci_2004}, and convolutional neural networks (CNN) \cite{cecotti_convolutional_2011}---typically operate on pre-defined features and do not provide an explicit mechanism for identifying which time points are most informative. Second, a line of work combines pre-processing or feature selection with relatively simple classifiers, such as genetic-algorithm-based selection with SVM \cite{schroder_automated_2003}, PCA and forward stepwise selection with neural networks \cite{rejer_eeg_2012}, common spatial patterns (CSP) with LDA \cite{bhatti_soft_2019}, and stepwise LDA (swLDA) using p-value-based inclusion/exclusion rules \cite{donchin_mental_2000, krusienski_toward_2008}; while often effective, these pipelines generally lack a unified probabilistic interpretation of temporal feature relevance. Third, recent Bayesian frameworks introduce more structured and interpretable feature selection for ERP-based BCIs, including the split-and-merge Gaussian process (SMGP) model for sequence-level discrepancies \cite{ma_bayesian_2022} and the GLASS model with sparse time-varying effects \cite{zhao_bayesian_2025}. However, SMGP is computationally intensive under the sequence-based framework, and GLASS requires additional data arrangement to split each full sequence of data into two half sequences to facilitate modeling.

To address the limitation of existing methods, building upon the SMGP-based framework of \citet{ma_bayesian_2022}, we propose a Bayesian classification framework with a Probit-link Split-and-Merge Gaussian Process (P-SMGP) prior for feature selection and classification in P300 ERP-based BCI speller system. Our method differs from the existing methods in the following ways: (1) We shift from sequence-based modeling to stimulus-level modeling to facilitate binary classification, and the feature selection provides a direct interpretation of separation effects between target and non-target ERPs without additional data arrangement. (2) 
\textcolor{violet}{We adopt a probit link instead of the truncated normal formulation to simplify posterior inference;}
(3) We apply a weighted-likelihood-based approach to update the character-level probability instead of direct summing up the classifier scores to leverage the potential dependence relationship among stimuli within one sequence.

The remaining paper is organized as follows: Sections 2 and 3 detail the formulation of the Bayesian generative model and its posterior inference framework, respectively. Sections 4 and 5 present the model setups and results from simulation studies and real-data analysis, respectively. Finally, Section 6 concludes with a brief discussion.

\section{Bayesian Modeling of EEG-BCI Data}

\subsection{Problem Setup and Notations}

Let $\mathcal{N}(\mu,\sigma^2)$ be a normal distribution with mean $\mu$ and variance $\sigma^2$. Let $\mathcal{MN}(\mu,\Sigma)$ be a multivariate normal distribution with mean $\mu$ and variance matrix $\Sigma$. Let $\mathcal{MN}(M,U,V)$ be a matrix normal distribution with location matrix $M$ and two scale matrix parameters $U$ and $V$. Let $\mathcal{GP}(\mu,\kappa)$ be a Gaussian process with mean function $\mu$ and kernel function $\kappa$. Let $\mathcal{U}(a,b)$ be a Uniform distribution with lower and upper bounds $a$ and $b$. Let $\mathcal{HC}(\sigma,\lambda)$ be a half-Cauchy distribution with location parameter $\sigma$ and scale parameter $\lambda$. Let $\mathcal{TN}_{[a,b]}(\mu, \sigma^{2})$ be a truncated normal distribution with mean $\mu$, variance $\sigma^{2}$, and truncation bounds $a$ and $b$. Let $\mathcal{LN}(\mu, \sigma^{2})$ be a log-normal distribution with location parameter $\mu$ and scale parameter $\sigma^{2}$.

Our model emphasizes the multi-channel EEG data of a single participant. Let $l(l=1,...,L)$ and $i(i=1,...,I)$ be the participant's character index and sequence index, respectively. We adhere to the standard RCP design, where every sequence comprises $J(J=12)$ stimuli. This set includes six stimuli arranged in rows $(1,...,6)$ and six in columns $(7,...,12)$ on the virtual keyboard grid of size $6 \times 6$. 
\textcolor{violet}{For the \(i\)th sequence of the \(l\)th target character, let \(W_{l,i}=(W_{l,i,1},\ldots,W_{l,i,12})^\top\) denote the stimulus-order vector, where \(W_{l,i,j}\) records the identity of the stimulus presented at the \(j\)th position in the sequence. The vector \(W_{l,i}\) is a permutation of \(\{1,\ldots,12\}\).  Equivalently, if \(W_{l,i,j}=w\leq 6\), the \(j\)th stimulus is row \(w\); if \(W_{l,i,j}=w>6\), the \(j\)th stimulus is column \(w-6\). For example, \(W_{l,i}=(9,1,4,8,6,2,7,10,12,3,11,5)^\top\) indicates that the first stimulus is column 3, the second stimulus is row 1, the third stimulus is row 4, and so on.}
 Let $Y_{l,i}=(Y_{l,i,1},...,Y_{l,i,12})^\top$ represent a stimulus-type indicator, where $Y_{l,i,j}\in\{0,1\}$ with the constraint $\sum_{j=1}^{6}Y_{l,i,j}=\sum_{j=7}^{12}Y_{l,i,j}=1$. For example, $Y_{l,i}=(0,1,0,0,0,0,0,0,0,1,0,0)^\top$ indicates that the target letter is located at the second row and the fourth column. We collect EEG segments starting from the onset of each stimulus, using an extended EEG response window of duration $T$. Finally, we examine $E$ channels of EEG signals and let $e(e = 1, ..., E)$ index the channel, and we denote $X_{l,i,j,e}(t)$ as the observed EEG signal segment of the $l$th target character, $i$th sequence and $j$th stimulus from channel $e$ at time $t\in[0, T]$.
For simplicity, we assume that participants are spelling the same character so that we remove participant’s character index $l$ in the following content.

\subsection{A Bayesian Model}

For a single participant and $\forall t\in[0, T]$, we assume 
\begin{equation}\label{1}
\begin{split}
\bm{X}_{i,j,e}(t)&=\bm{M}_{i,j,e}(t)+\bm{\epsilon}_{i,j,e}(t),\\
\bm{M}_{i,j,e}(t)&=\bm{\beta}_{1,e}(t)Y_{i,j}+\bm{\beta}_{0,e}(t)(1-Y_{i,j}),
\end{split}
\end{equation}
where $\bm{\beta}_{1,e}(t)$ and $\bm{\beta}_{0,e}(t)$ are the average brain activity responses to the target and the non-target stimulus, respectively. For simplicity, we assume that the shape and magnitude of ERP functions depend solely on the stimulus-type indicators, irrespective of the stimulus location or order.

For random noise  $\epsilon_{i,j,e}(t)$, we consider an additive model for every discrete time point $t$ as follows:
\begin{align*}
\epsilon_{i,j,e,t} &= \xi_{i,j,e} + \varepsilon_{i,j,t},\\
\bm{\xi}_{i,j} &= (\xi_{i,j,1},...,\xi_{i,j,E})^\top \sim \mathcal{MVN}(0,C_{s}),\\
\varepsilon_{i,j,t} &= \rho_{t}\varepsilon_{i,j,t-1} + \eta_{i,j,t},\\
\varepsilon_{i,j,1}&\sim \mathcal{N}(0,\sigma^{2}_{\rho_{t}}),\quad \eta_{i,j,t} \sim \mathcal{N}(0,\sigma^{2}_{\rho_{t}}),
\end{align*}
where $\bm{\xi}_{i,j}$ is a vector of channel-specific random effect which follows a multivariate normal distribution with the mean 0 and the covariance matrix $C_{s}$ that has a compound symmetry structure and $\varepsilon_{i,j,t}$ is temporal-specific random effect which is assumed to follow an autoregressive  model of first order with coefficient $\rho_{t}$ and noise variance $\sigma^{2}_{\rho_{t}}$ \cite{ma_bayesian_2026}.

\subsection{The Probit-link Split-and-Merge GP}
We apply the Probit-link Split-and-Merge GP (P-SMGP) prior to model the joint prior distribution of $\bm{\beta}_{1,e}(t)$ and $\bm{\beta}_{0,e}(t)$, for $t\in[0, T]$. We assume $\bm{\beta}_{k,e}(t)(e=1,...,E)$ are independent, and each marginal distribution adheres to the prior distribution specified by the SMGP, where $k\in\{0,1\}$. For simplicity, we omit the channel-specific index $e$ when defining the SMGP as follows:
\begin{equation}\label{2}
\bm{\beta}_{k}(t)=\bm{\alpha}_{k}(t)\bm{\zeta}(t)+\bm{\alpha}_{0}(t)(1-\bm{\zeta}(t)),
\end{equation}

\textcolor{violet}{where $\bm{\alpha}_{k}(t)\sim\mathcal{GP}(0,\kappa_{\alpha})$ and the split-and-merge weight $\bm{\zeta}(t)\in[0,1]$ is defined through a probit-link latent Gaussian process:
\begin{align}
\bm{\zeta}(t)
&=
\Phi\{\bm{\omega}(t)\},\\
\bm{\omega}(t)
&\sim
\mathcal{GP}\left(0,\psi_{\omega}\kappa_{C_{\omega}}\right),
\end{align}
where $\Phi(\cdot)$ denotes the standard normal cumulative distribution function. The covariance kernel of the latent process is specified as
\begin{equation}
\kappa_{C_{\omega}}(t, t')
=
\begin{cases}
1, & t=t',\\[3pt]
\rho_{\omega},
& 0<|t-t'|\leq\delta_{\omega},\\[3pt]
0, & |t-t'|>\delta_{\omega},
\end{cases}
\label{eq:omega_kernel}
\end{equation}
where $\psi_{\omega}>0$ is the marginal variance, $\rho_{\omega}$ controls the correlation between neighboring time points, and $\delta_{\omega}$ is a distance threshold. In our analysis, $\delta_{\omega}$ is set equal to the temporal distance between two adjacent EEG sampling points. This locally supported covariance kernel reflects the assumption that the tendency of the signal to split or merge at a given time point is associated only with the immediately adjacent time points. Time points separated by more than one sampling interval are assumed to be conditionally unrelated in the latent split-and-merge process.}

\begin{figure}[htbp]
  \centering

  \begin{minipage}[b]{0.31\textwidth}
    \centering
    \includegraphics[width=\textwidth]{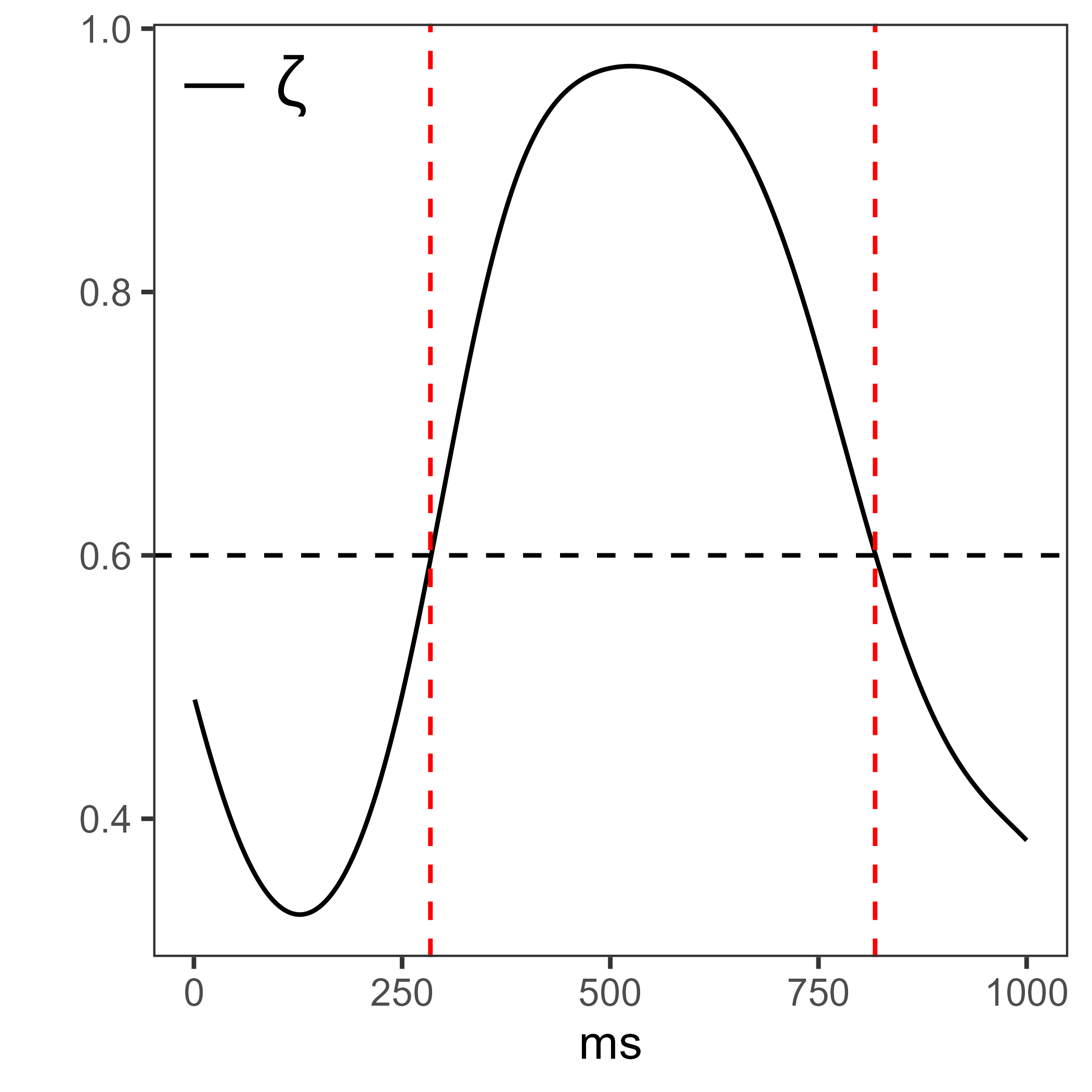}
    {\small (a)}
  \end{minipage}
  \hfill
  \begin{minipage}[b]{0.31\textwidth}
    \centering
    \includegraphics[width=\textwidth]{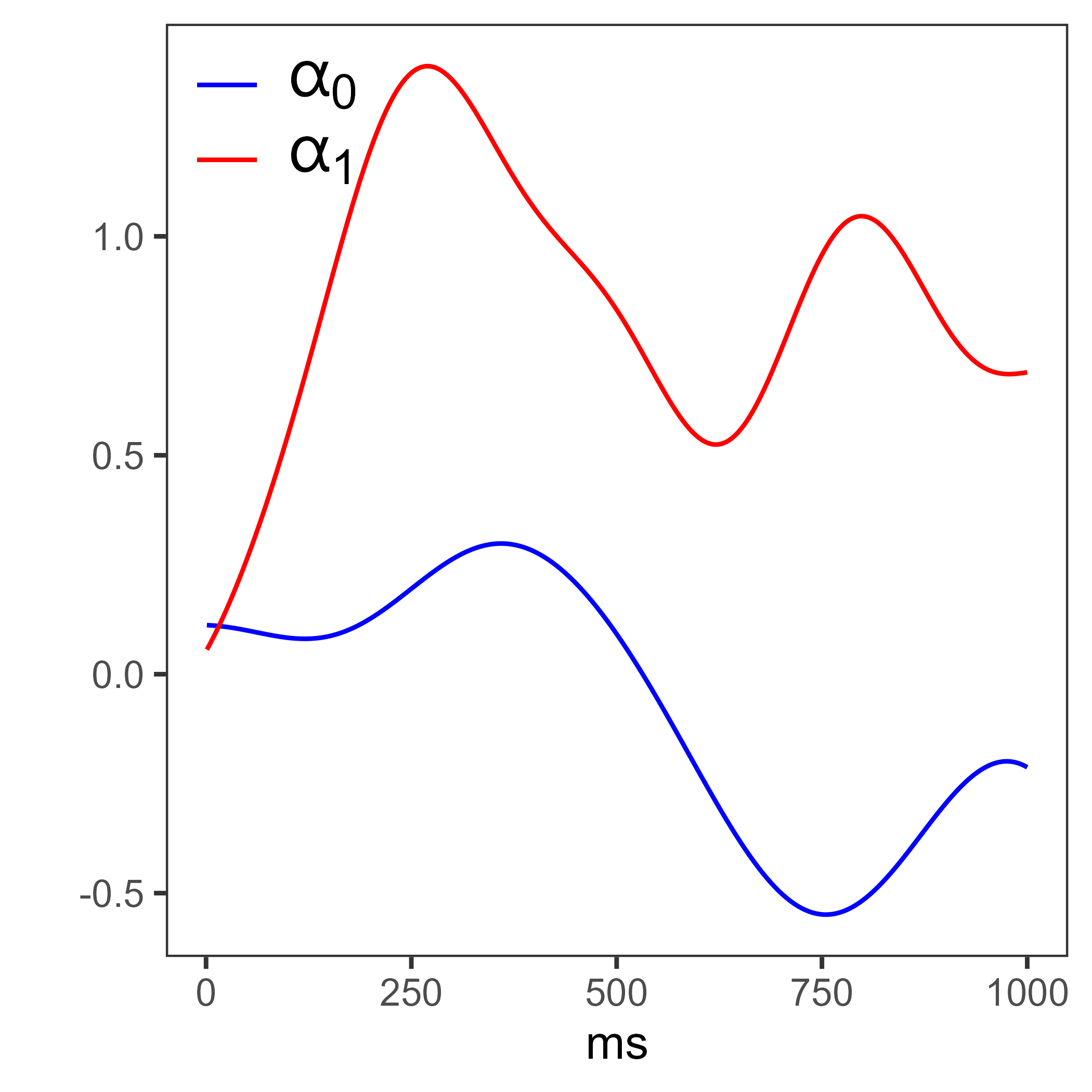}
    {\small (b)}
  \end{minipage}
  \hfill
  \begin{minipage}[b]{0.31\textwidth}
    \centering
    \includegraphics[width=\textwidth]{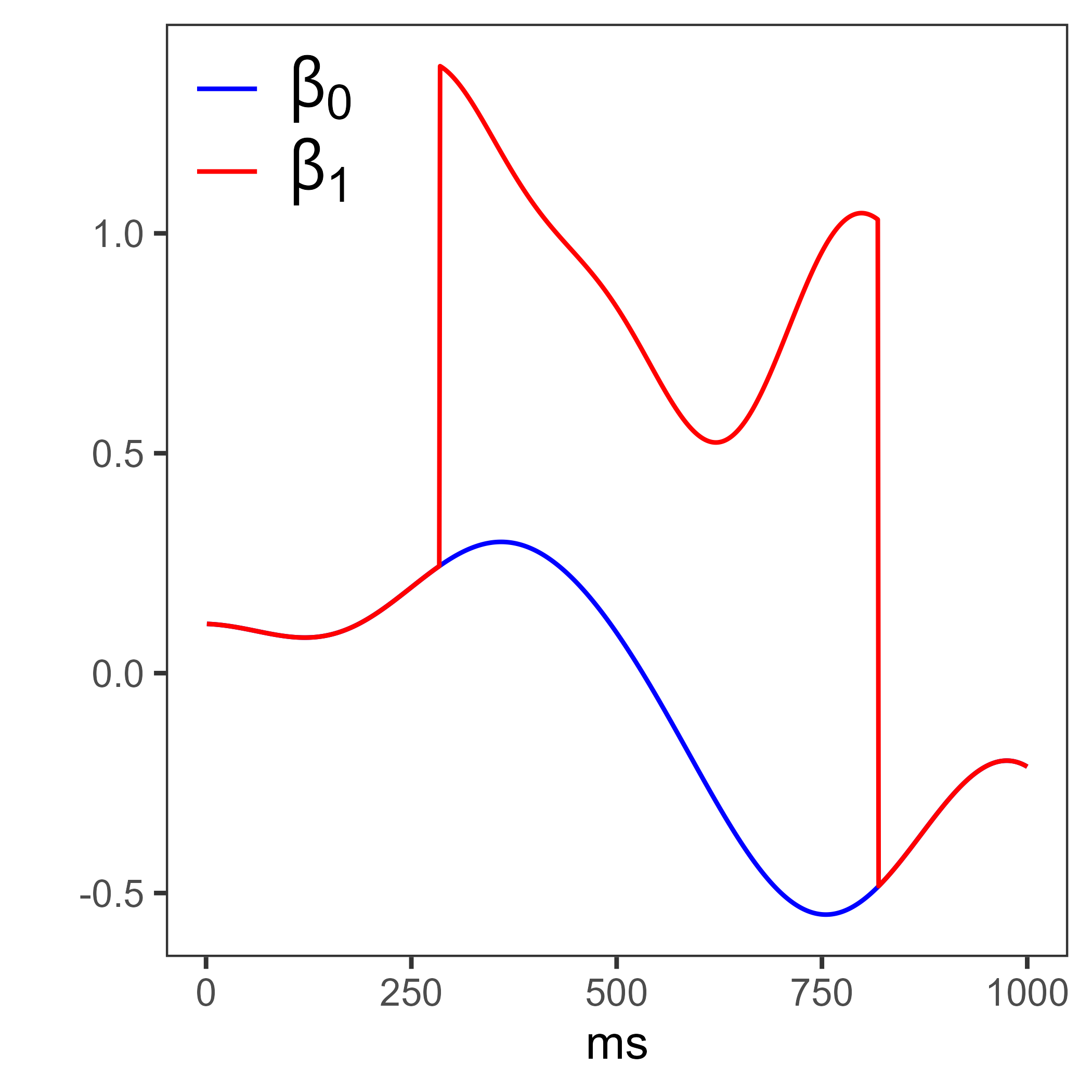}
    {\small (c)}
  \end{minipage}

  \caption{\small (a) The 
  \textcolor{violet}{split-and-merge weight}
  $\zeta$ with 0.6 as the threshold. (b) Two simple Gaussian processes $\alpha_{1}$ and $\alpha_{0}$ before the split-and-merge process. (c) Two Gaussian processes $\beta_{1}$ and $\beta_{0}$ after the split-and-merge process based on $\zeta$.}
  \label{fig:3}
\end{figure}

\textcolor{violet}{Let $\{t_1,\ldots,t_T\}$ denote the equally spaced empirically observed times. Evaluating $\kappa_{C_{\omega}}$ over this grid yields a covariance matrix of 
\begin{equation}
\bm{C}_{\omega}
=
\psi_{\omega}
\begin{pmatrix}
1 & \rho_{\omega} & 0 & \cdots & 0\\
\rho_{\omega} & 1 & \rho_{\omega} & \ddots & \vdots\\
0 & \rho_{\omega} & 1 & \ddots & 0\\
\vdots & \ddots & \ddots & \ddots & \rho_{\omega}\\
0 & \cdots & 0 & \rho_{\omega} & 1
\end{pmatrix}.
\end{equation}
}
The 
\textcolor{violet}{split-and-merge weight}
process is denoted as $\bm{\zeta}(t)$ because $\bm{\beta}_{0}(t)=\bm{\alpha}_{0}(t)$ and 
$\bm{\beta}_{1}(t)$ is the weighted average between $\bm{\alpha}_{0}(t)$ and $\bm{\alpha}_{1}(t)$ by $\bm{\zeta}(t)$. 
\textcolor{violet}{After posterior estimation, we apply a hard-threshold selection method in the split-and-merge process, controlled by a hyperparameter $\zeta_{0}$.}
We will discuss the rationale of selecting $\zeta_0$ later. Specifically, when $\bm{\zeta}(t) \leq \zeta_{0}$, $\bm{\beta}_{1}(t)$ is set equal to $\bm{\beta}_{0}(t)$, indicating no separation. However, when $\bm{\zeta}(t)>\zeta_{0}$, $\bm{\beta}_{1}(t)$ and $\bm{\beta}_{0}(t)$ are split, allowing them to take on different values. Let $\mathcal{W}_{s}=\{t:\bm{\zeta}(t)>\zeta_{0}\}$ and $\mathcal{W}_{m}=\{t:\bm{\zeta}(t) \leq \zeta_{0}\}$ represent the split time window and the merge time interval, respectively. Figure \ref{fig:3} shows the process in a more visual way.

\section{Posterior Inferences}
\subsection{Model Representation and Prior Specification}
We rewrite Equation (\ref{1}) in the form of matrix normal distribution
\begin{equation}\label{3}
\bm{X}_{i,j} \sim \mathcal{MN}(\bm{M}_{i,j},C_{t},C_{s}),
\end{equation}
where $\bm{X}_{i,j}=(\bm{X}_{i,j,e})_{e=1}^{E}$ and $\bm{M}_{i,j}=(\bm{M}_{i,j,e})_{e=1}^{E}$ are matrix-wise observed EEG signals and predicted EEG signals for the $i$th sequence and $j$th stimulus, respectively. $C_{s}$ and $C_{t}$ are the channel-specific and temporal-specific covariance matrices jointly characterizing the random error $\bm{\epsilon}_{i,j}=(\bm{\epsilon}_{i,j,e})_{e=1}^{E}$, respectively.

In the P-SMGP process, $\bm{\alpha}_{0,e}$ and $\bm{\alpha}_{1,e}$ are channel-specific response functions to non-target and target stimuli before selection. We assume that they follow the $\mathcal{GP}(0,\kappa_{\alpha})$ with the scale parameters $\psi_{0,e}$ and $\psi_{1,e}$, respectively. We use a $\gamma$-exponential function to specify non-target kernel $\kappa_0$ and target kernel $\kappa_1$ as follows:
\begin{align*}
&k_{e}(x_i, x_j) = \psi_{e} \exp \left\{ -\left( \frac{\|x_i - x_j\|^2}{s} \right)^{\gamma} \right\},
\end{align*}
where $0 \leq{\gamma} < 2, {s} > 0$. In practice, $\gamma$ and $s$ are treated as hyper-parameters. We systematically select their values by first generating prior functions from a wide range of combinations $(\gamma, s)$, then identifying the optimal parameters whose corresponding functions exhibit morphological similarity to empirical ERP waveforms. 

In addition, for a kernel function $k(x, x')$ with eigenvalue $\lambda$, the eigenfunction can be defined as $\varphi(\cdot)$ if this integral relationship holds:
\begin{align*}
\int k(x, x')\varphi(x)dx = \lambda \varphi(x')
\end{align*}
The set of eigenfunctions, $\{\varphi_l(x)\}_{l=1}^{\infty}$, is generally infinite, and we order them in descending magnitude of their associated eigenvalues, $\{\lambda_l\}_{l=1}^{\infty}$. According to Mercer's Theorem, the kernel function $k(x, x')$ can be expressed as a weighted sum of normalized eigenfunctions, where each eigenfunction $\varphi_l(s)$ is paired with a positive eigenvalue $\lambda_l$. This decomposition allows for an approximation of $k(x, x')$ using the first $L$ terms of the series, provided $L$ is sufficiently large, as follows:
\[
k(x, x') = \sum_{l=1}^{+\infty} \lambda_l \varphi_l(x)\varphi_l(x') \approx \sum_{l=1}^{L} \lambda_l \varphi_l(x)\varphi_l(x').
\]

Therefore, by Mercer’s representation theorem, we obtain that
\begin{align*}
\bm{\alpha}_{k}(t)&=\psi_{k}\sum_{l=1}^{+\infty}\text{Diag}(\bm{\theta}_{k,l})\varphi_{k,l}(t)\approx
\psi_{k}\sum_{l=1}^{L_{k}}\text{Diag}(\bm{\theta}_{k,l})\varphi_{k,l}(t), \quad k=0,1
\end{align*}
where $\bm{\theta}_{k,l}$ are column vectors of coefficient parameters associated with eigenfunctions $\varphi_{k,l}(t)$. We evaluate $\bm{\alpha}_{1}(t)$ and $\bm{\alpha}_{0}(t)$ over a group of input time points $\{t_{n}\}_{n=1}^{T_{0}}$ and define them $\bm{A}_{1}$ and $\bm{A}_{0}$, respectively. Then, we adopt the kernel function $\kappa_{1}$ and $\kappa_{0}$ to calculate the kernel covariances $\bm{\Psi}_{1}$ and $\bm{\Psi}_{0}$, respectively. $\bm{A}_{1}$ and $\bm{A}_{0}$ can be approximated by
\begin{align*}
&\underset{T_{0} \times 1}{\bm{A}_{1}}\approx\psi_{1}\underset{T_{0} \times L_{1}}{\bm{\Psi}_{1}^{1:L_{1}}} \underset{L_{1} \times 1}{\bm{\theta}_{1}},\quad \bm{\theta}_{1}=(\theta_{1,1}\ldots\theta_{1,l}\ldots\theta_{1,L_{1}}),\\
&\underset{T_{0} \times 1}{\bm{A}_{0}}\approx\psi_{0}\underset{T_{0} \times L_{0}}{\bm{\Psi}_{0}^{1:L_{0}}} \underset{L_{0} \times 1}{\bm{\theta}_{0}},\quad \bm{\theta}_{0}=(\theta_{0,1}\ldots\theta_{0,l}\ldots\theta_{0,L_{0}}),\\
&\bm{\theta_{1}}\sim\mathcal{MVN}(\mathbf{0},\bm{V}_{\bm{\Psi}_{1}}^{1:L_{1}}),\quad \bm{\theta_{0}}\sim\mathcal{MVN}(\mathbf{0},\bm{V}_{\bm{\Psi}_{0}}^{1:L_{0}}),
\end{align*}

where $\bm{\Psi}_{1}^{1:L_{1}}$ , $\bm{\Psi}_{0}^{1:L_{0}}$ , $\bm{V}_{\bm{\Psi}_{1}}^{1:L_{1}}$ , and $\bm{V}_{\bm{\Psi}_{0}}^{1:L_{0}}$ are the first $L_{1}$ rows of $\bm{\Psi}_{1}$, the first $L_{0}$ rows of $\bm{\Psi}_{0}$, the diagonal matrix of the first $L_{1}$ eigenvalues associated with $\bm{\Psi}_{1}$, and the diagonal matrix of the first $L_{0}$ eigenvalues associated with $\bm{\Psi}_{0}$, respectively \cite{zhang_highdimensional_2021}. In practice, $L_{1}$ and $L_{0}$ are determined by the minimum integer for which the cumulative
sum of eigenvalues divided by the total sum of eigenvalues is over 95\% after reordering them by the descending order. Finally, the prior specifications are as follows:
\begin{align*}
&\bm{\theta_{1}}\sim\mathcal{MVN}(\mathbf{0},\bm{V}_{\bm{\Psi}_{1}}^{1:L_{1}}),\quad \bm{\theta_{0}}\sim\mathcal{MVN}(\mathbf{0},\bm{V}_{\bm{\Psi}_{0}}^{1:L_{0}}),\\
& \bm{\omega}_{e}\sim\mathcal{MVN}(0,\psi_{\omega}C_{\omega}),\quad \bm{\zeta}_{e,t}=\Phi(\bm{\omega}_{e,t}), \quad \mathrm{elementwise},\\
&\sigma^{2}_{\rho_{t}}\sim\mathcal{HC}(0, 5.0),\quad \rho_{t}\sim\mathcal{TN}_{[0,1]}(0.5,1),\\
&\psi_{1}, \psi_{0} \overset{\text{indep.}}{\sim} \mathcal{LN}(0, 1).
\end{align*}

\subsection{Markov Chain Monte Carlo}
The Bayesian inference was implemented using Markov Chain Monte Carlo (MCMC) methods through the Python package \texttt{NumPyro} \cite{phan_composable_2019,bingham_pyro_2019}, which utilizes \texttt{JAX}-based just-in-time compilation to facilitate posterior computation. We employed the No-U-Turn Sampler (NUTS) \cite{homan_no-u-turn_2014}, an adaptive variant of Hamiltonian Monte Carlo, to efficiently sample from potentially complex posterior distributions. For the convergence check, we run multiple chains with different seed values and evaluate the convergence by the Gelman-Rubin statistic \cite{gelman_inference_1992}. The posterior samples obtained were then used for parameter estimation, ensuring that the results are both accurate and computationally efficient.

\subsection{Posterior Character-Level Probability for Prediction}
Under the RCP design, the selection of the target character requires the selection of both a target row and a target column, each chosen from a respective pool of six candidate rows and six candidate columns. 
We denote each candidate character by $l \in \mathcal{L}$, where $|\mathcal{L}| = 36$. 
Each character is uniquely identified by a row index $r(l) \in \{1,\dots,6\}$ and a column index $c(l) \in \{7,\dots,12\}$. For the $i$ th sequence ($i = 1,2,\dots$) and the $j$ th flash, we denote the flashed row/column by $F_{i,j} \in \{1,\dots,12\}$ and the observed EEG segment by $\bm{X}_{i,j}$. Based on the matrix normal model, the corresponding log-likelihood is defined as
\[
s_y(i,j) = \log p(\bm{X}_{i,j} \mid Y_{i,j} = y),\quad y=0,1.
\]

If the target character is $l$, then in sequence $i$, flashes corresponding to $r(l)$ or $c(l)$ are treated as targets ($Y=1$), while the others are non-targets ($Y=0$). 
The log-likelihood for $l$ in $i$-th sequence is therefore:
\[
\mathrm{Log}L_i(l) = \sum_{j=1}^{12} \Big[ 
\mathbf{1}\{F_{i,j} \in \{r(l), c(l)\}\}\, s_1(i,j) 
+ \mathbf{1}\{F_{i,j} \notin \{r(l), c(l)\}\}\, s_0(i,j) 
\Big].
\]

After the first $i$ sequences, the cumulative log-likelihood for character $l$ is
\[
\mathrm{Cum Log}L_i(l) = \sum_{k=1}^{i} L_k(l).
\]

Let $\pi_0(l)$ denote the prior distribution over characters (typically uniform when no prior information exists about the target character, $\pi_0(l)=1/36$). 
The posterior probability that character $l$ is the target after observing the first $i$ sequences is obtained by Bayes’ rule \cite{throckmortonBayesianApproachDynamically2013}:
\[
\pi_i(l) = 
\frac{\pi_0(l) \, \exp\{\mathrm{Cum Log}L_i(l)\}}
{\sum_{l' \in \mathcal{L}} \pi_0(l') \, \exp\{\mathrm{Cum Log}L_i(l')\}}.
\]

Thus, the probability of $l$ being the true target character is updated after each sequence cumulatively.

\section{Simulation Studies}
We have performed several multi-channel simulation studies and compared BLDA and swLDA as reference methods to demonstrate the advantages of our proposed method.

\subsection{Setup}
We randomly generated stimulus-type indicators with 20 characters for our simulation. The simulated EEG signals were generated with 2 channels, i.e., $E=2$, and a response window of 30 time points per channel, i.e., $T=30$. For both target and non-target EEG signals, their shapes and magnitude were based on real participants from an existing database \cite{thompson_plug-and-play_2014}, and we consider the same variance structure for them, which is an autoregressive temporal structure of order 1 (i.e., AR(1)), compound symmetry spatial structure, and channel-specific variances for background noises, where the true parameters were $\sigma_{\rho_{t}} = 2, \rho_{t} = 0.6, \rho_{e} = 0.6$.  We performed 100 dataset replications for this scenario. For each dataset, we generated 10 and 15 sequences per character for training and testing, respectively. These parameters are determined based on the real data analysis.

\subsection{Model Fitting and Diagnostics}
All simulated datasets were fitted with the model described in Equation (\ref{1}). We also applied two covariance kernels $\kappa_1$ and $\kappa_0$ for target and non-target ERP functions, respectively. The two kernels were assumed with a $\gamma$-exponential kernel, and the length-scale, gamma and scaling parameters of target and non-target stimuli were ($s_1 = 6.5, \gamma_1 = 1.5, \psi_1 = 1$) and ($s_0 = 9.5, \gamma_0 = 1.5, \psi_0 = 1$), respectively. Additionally, we consider the same P-SMGP process for both target and non-target EEG signals, where the true parameters were $\psi_{\omega} = 40, \rho_{\omega} = 0.5$. The pre-specified threshold $\zeta_0$ was set to 0.8 because the simulated signal function had relatively small variation and showed clear time-dependent patterns. This led us to use a higher threshold --- the split would only be triggered when differences between regions were large enough to clearly stand out from random background variations.. We ran the MCMC with 2 chains, with each chain containing 2,000 burn-ins and 1,000 MCMC samples. The MCMC convergence was assessed by running different chains with different seeds and the Gelman-Rubin statistic.

\subsection{Results}
\subsubsection{Parameter Estimation}
\begin{figure}[htbp]
  \centering

  \begin{minipage}[c]{0.45\textwidth}
    \centering
    \includegraphics[width=\linewidth]{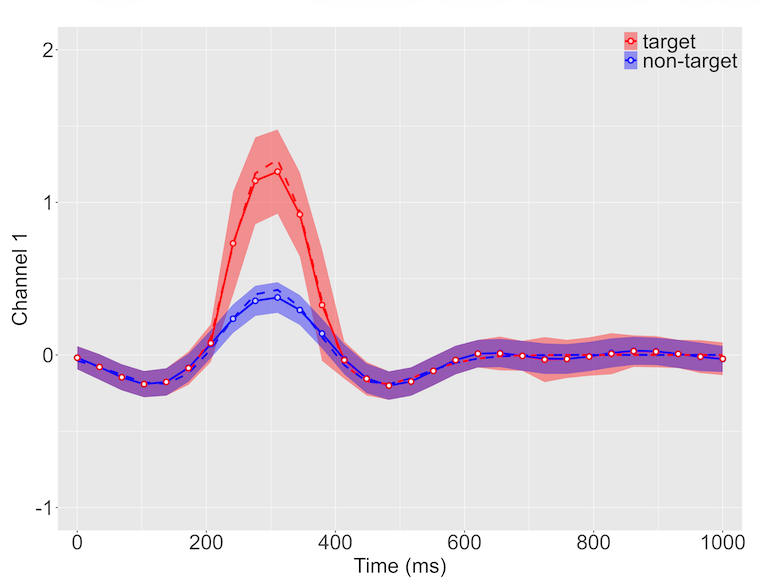}
    {\small (a)}
  \end{minipage}
  \hfill
  \begin{minipage}[c]{0.45\textwidth}
    \centering
    \includegraphics[width=\linewidth]{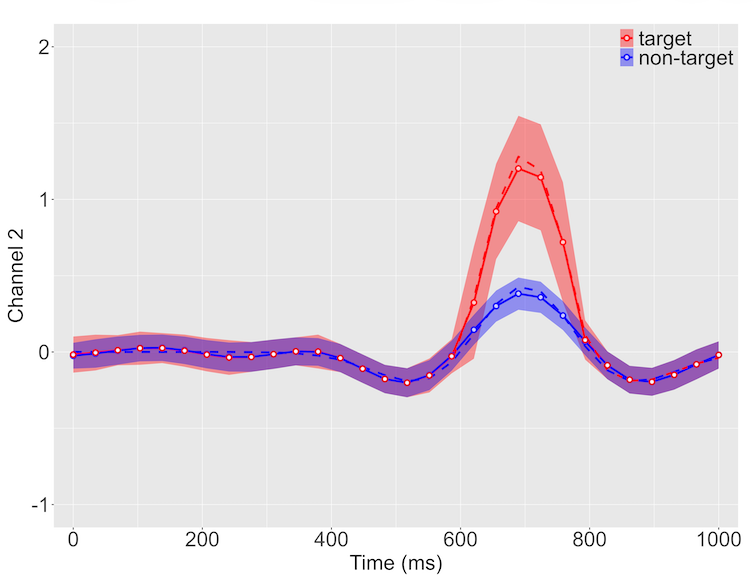}
    {\small (b)}
  \end{minipage}


  \begin{minipage}[c]{0.45\textwidth}
    \centering
    \includegraphics[width=\linewidth]{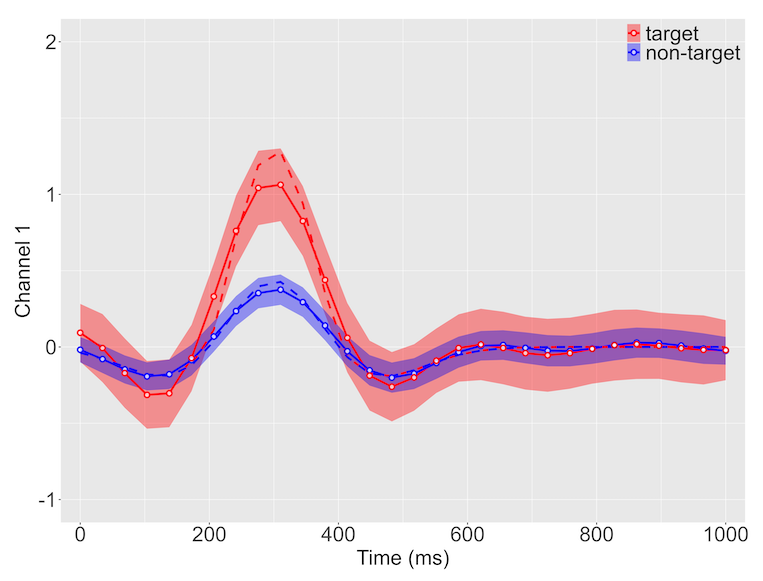}
    {\small (c)}
  \end{minipage}
  \hfill
  \begin{minipage}[c]{0.45\textwidth}
    \centering
    \includegraphics[width=\linewidth]{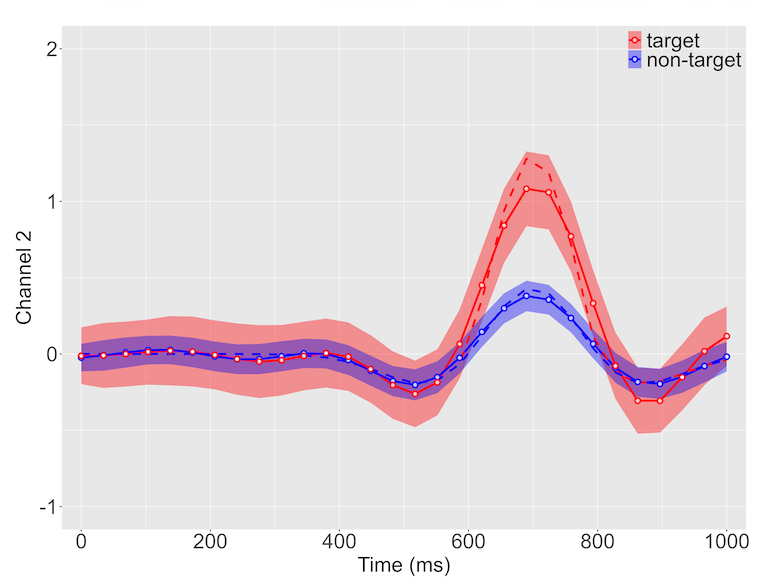}
    {\small (d)}
  \end{minipage}
  \caption{\small The estimated target and non-target ERP functions for the simulated dataset are shown in two rows: the upper row (a, b) uses P-SMGP, and the lower row (c, d) uses BLDA. Panels (a) and (c) display results for Channel 1, while panels (b) and (d) display results for Channel 2. The target ERP functions are shown in red, and the non-target ERP functions are shown in blue.}
  \label{fig:4}
\end{figure}

To evaluate the estimation performance of our proposed method, we compared the estimated ERP functions ($\bm{\beta}_{1}, \bm{\beta}_{0}$) with their true values for each time point in the two EEG channels. Figure \ref{fig:4} (a) and Figure \ref{fig:4} (b) present the results for the P-SMGP method, while Figure \ref{fig:4} (c) and Figure \ref{fig:4} (d) correspond to the BLDA method. Since swLDA is a discriminant method instead of a generative method, it cannot provide  parameter estimates. Each sub-figure of Figure 4 illustrates the posterior mean estimates of $\bm{\beta}_{1}$ and $\bm{\beta}_{0}$ along with their corresponding 95\% credible intervals, compared to the ground-truth values. The red and blue solid lines represent the posterior mean estimates of $\bm{\beta}_{1}$ and $\bm{\beta}_{0}$, respectively, while the dashed lines indicate the true values. The shaded regions indicate the 95\% credible intervals.

The P-SMGP and BLDA approach both provide accurate estimates for both $\bm{\beta}_{1}$ and $\bm{\beta}_{0}$, with the true values falling within the 95\% credible intervals across most time points. The peak responses at around 250 ms (Channel 1) and 750 ms (Channel 2) are well captured, demonstrating the models' ability to recover ERP components with high precision. Notably, for P-SMGP method, the credible intervals for $\bm{\beta}_{1}$ and $\bm{\beta}_{0}$ remain consistently similar in width in regions where the signal exhibits minimal variation. In contrast, the BLDA method, while also providing reasonable mean estimates, exhibits considerably wider credible intervals for $\bm{\beta}_{1}$ compared to $\bm{\beta}_{0}$, particularly in regions of lower signal magnitude.

\subsubsection{Prediction Performance}
To further evaluate the performance of different methods, we compared their character-level prediction accuracy across different sequence sizes. Table \ref{tab:1} presents the accuracy results for P-SMGP, swLDA, and BLDA. The testing prediction accuracy increased as the sequence size grows. Among the three methods, P-SMGP consistently achieved the highest accuracy across all sequence sizes. Notably, at relatively smaller sequence sizes (e.g., 6–10). As the sequence size increased, all methods converged towards high accuracy, but P-SMGP maintained a slight edge, reaching 90\% accuracy at a sequence size of 15, compared to 88\% for swLDA and 87\% for BLDA.

\begin{table}[ht]
\centering
\small 
\caption{Prediction accuracy of P-SMGP, swLDA and BLDA for simulation datasets with different sequence sizes (mean ± standard deviation).}
\label{tab:1}
\begin{tabularx}{\textwidth}{>{\centering\arraybackslash}X | *{3}{>{\centering\arraybackslash}X}} 

\hline
\multirow{2}{*}{Sequence Size} & \multicolumn{3}{c}{Methods}       \\ \cline{2-4} 
                               & P-SMGP      & swLDA     & BLDA      \\ \hline
6                              & 0.60±0.12 & 0.55±0.12 & 0.57±0.11 \\
7                              & 0.66±0.12 & 0.59±0.13 & 0.60±0.11 \\
8                              & 0.70±0.10 & 0.65±0.12 & 0.66±0.10 \\
9                              & 0.75±0.10 & 0.70±0.12 & 0.70±0.10 \\
10                             & 0.79±0.10 & 0.74±0.10 & 0.73±0.10 \\
11                             & 0.81±0.09 & 0.78±0.09 & 0.76±0.09 \\
12                             & 0.84±0.08 & 0.81±0.09 & 0.80±0.08 \\
13                             & 0.87±0.08 & 0.83±0.08 & 0.83±0.08 \\
14                             & 0.88±0.07 & 0.85±0.08 & 0.86±0.07 \\
15                             & 0.90±0.08 & 0.88±0.08 & 0.87±0.07 \\
\hline
\end{tabularx}
\end{table}

\section{Analysis of Real EEG-BCI Data}
We performed the analysis of EEG-BCI data, where the data were originally collected from the \if1\anon
{University of Michigan Direct Brain Interface (UMDBI) Laboratory}\fi
\if0\anon{XXX Lab}\fi. In total, 16 participants' data were used for model validation and reporting of results. The steps of the real data analysis were as follows: First, we preprocessed the raw EEG signals. Next, we fitted the model to the participants' data using the P-SMGP, BLDA, and swLDA methods, respectively, with the latter two methods considered to be the reference methods. Finally, we got the results of parameter estimation and classification accuracy. We presented detailed results for participant K178, an 56-year old white male with migraines, and presented results of the remaining participants in the Supplementary Material.

\subsection{Dataset and Pre-processing}
During the training session, each participant wore a 16-channel EEG cap 
and was seated approximately 0.8 meters from a 17-inch BCI display. Participants were instructed to type a 19-character test phrase, ``THE\_QUICK\_BROWN\_FOX,'' which included three spaces. Stimulus presentation and neural signal recording were implemented via the BCI2000 software platform \cite{schalk_bci2000_2004}. The study further required participants to perform tasks under three distinct interaction scenarios to evaluate the impact of environmental variations on BCI recognition accuracy: 1) direct input via the BCI interface (BCI type), 2) text entry using the standalone communication device DynaWrite (DYN type), and 3) typing on a laptop computer positioned adjacent to the BCI screen (CMP type). The sizes of super-sequences and sequences varied by participant. In the experimental design, an ``event'' was defined as the presentation of a single row or column stimulus, highlighted for 31.25 ms followed by a 125 ms inter-stimulus interval, forming a total stimulus-to-stimulus interval of 156.25 ms. A full cycle of all 12 row/column stimuli constituted a ``sequence,'' while multiple sequences were aggregated into a ``super-sequence.'' In our study, each super-sequence was associated with the EEG signals corresponding to one target character. During the training phase, data from 19 super-sequences were collected, each containing 15 repeated sequences. The sampling rate of the EEG device was 256 Hz.
For data pre-processing, first, raw EEG signals were preprocessed using a 0.5–6 Hz band-pass filter and down-sampled with a decimation factor of 8. Then, a fixed time window of 800 ms was extracted post stimulus to construct an EEG signal matrix per stimulus such that the raw truncated EEG signal segment has a dimension of (16, 25), where 16 and 25 refer to the number of channels and time points corresponding to about 800 ms post stimulus, respectively. To address the inherently low signal-to-noise ratio (SNR) of EEG signals, the xDAWN spatial filtering algorithm \cite{rivet_xdawn_2009} was applied. We selected the first two major components and defined the xDAWN-processed signals as ``transformed EEG signals.'' Each transformed EEG signal segment was a 50-dimensional vector.

\subsection{Model Setting}
The datasets were fitted with the model described in Equation \ref{1}.  For both target and non-target ERP functions, we employed two distinct covariance kernels, $\kappa_1$ (target) and $\kappa_0$ (non-target), each modeled as a $\gamma$-exponential kernel. Critically, the kernel hyper-parameters — including the length-scale, gamma, and scaling factor — were individually selected for each participant to account for inter-subject variability in neural dynamics. Additionally, we consider the same P-SMGP process for both target and non-target EEG signals, where the hyperparameters were $\psi_{\omega} = 40, \rho_{\omega} = 0.5$. The pre-specified threshold $\zeta_0$ was set to be 0.5 because actual neural signals had very low signal-to-noise ratios (SNR) --- by using a lower threshold, the system could retain relatively more differences between target and non-target signal components while still filtering out meaningless noise fluctuations. 

\textcolor{violet}{For participant K178, we adopted four independent NUTS chains, with 2,000 warm-up iterations and 1,000 retained posterior draws per chain. 
Convergence was evaluated using rank-normalized split $\widehat{R}$, bulk and tail effective sample sizes (ESS), divergent transitions, and representative trace and rank plots. For the key covariance and scale parameters and representative values of the latent ERP functions and split-and-merge weights, all $\widehat{R}$ values were below 1.01. The minimum bulk and tail ESS values were 859.7 and 900.9, respectively, and no divergent transitions were observed. Detailed diagnostic results are provided in Section S4.}

\subsection{Results}
\subsubsection{Parameter Estimation}
\begin{figure}[htbp] 
  \centering 
  \includegraphics[width=0.8\textwidth]{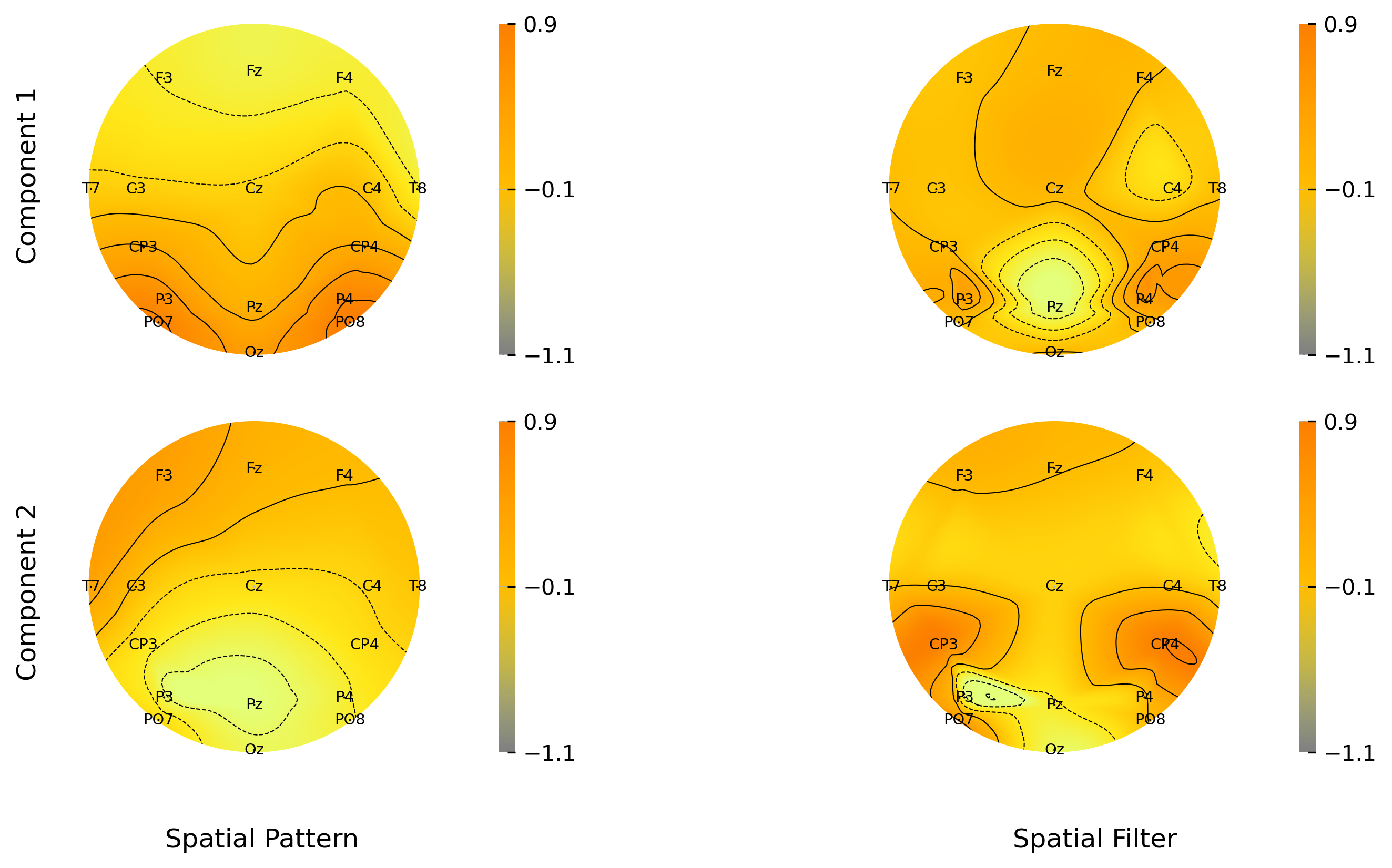} 
  \caption{\small The left and right columns show the spatial patterns and spatial filters of the participant K178's training data by xDAWN, respectively.} 
  \label{fig:6} 
\end{figure}

\begin{figure}[htbp]
  \centering

  \begin{minipage}[c]{0.45\textwidth}
    \centering
    \includegraphics[width=\linewidth]{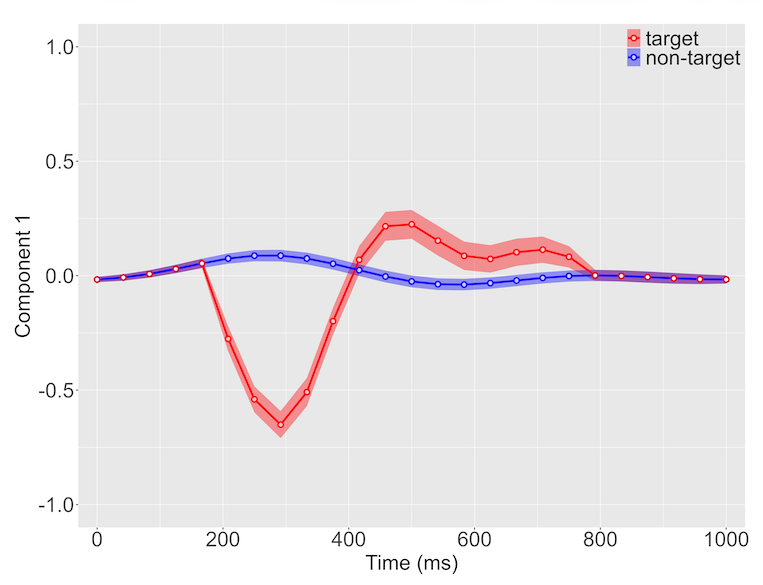}
    {\small (a)}
  \end{minipage}
  \hfill
  \begin{minipage}[c]{0.45\textwidth}
    \centering
    \includegraphics[width=\linewidth]{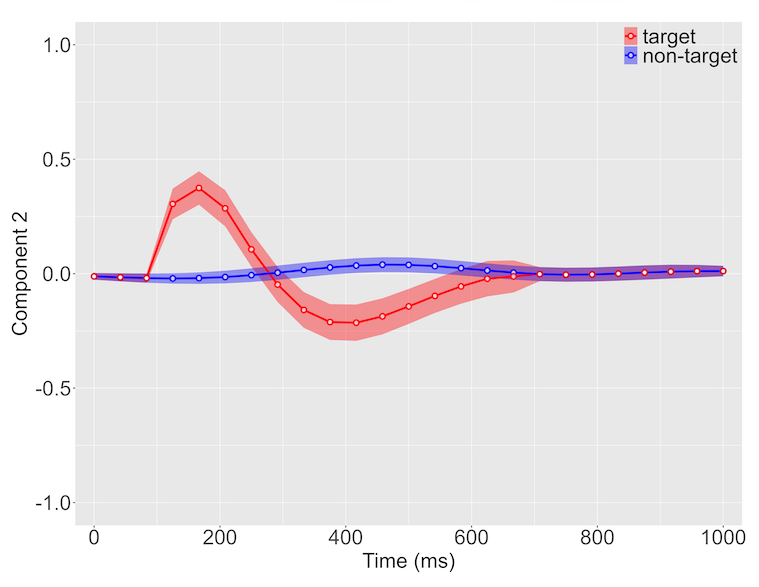}
    {\small (b)}
  \end{minipage}

  \vspace{1em}

  \begin{minipage}[c]{0.45\textwidth}
    \centering
    \includegraphics[width=\linewidth]{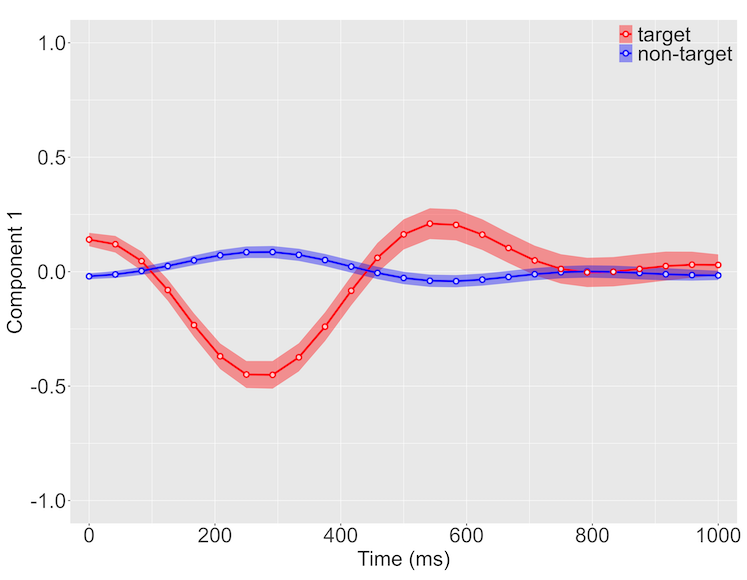}
    {\small (c)}
  \end{minipage}
  \hfill
  \begin{minipage}[c]{0.45\textwidth}
    \centering
    \includegraphics[width=\linewidth]{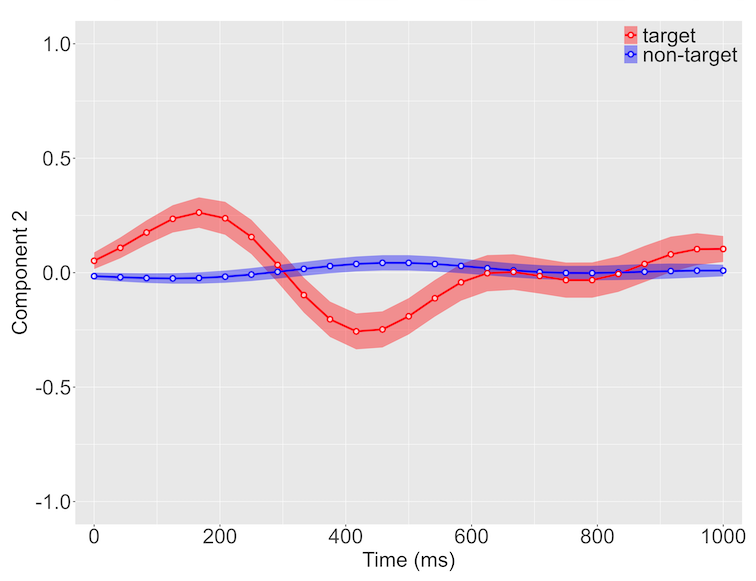}
    {\small (d)}
  \end{minipage}

  \caption{\small The estimated target and non-target transformed ERP functions of K178 by P-SMGP (a-b), and by BLDA (c-d). The left and right columns show the results of Components 1 and 2, respectively.}
  \label{fig:5}
\end{figure}

To assess the parameter estimation performance in the real data analysis for K178 within the BCI scenario, we examined the estimated ERP components ($\bm{\beta}_{1}$, $\bm{\beta}_{0}$) across time for the two extracted dimensions. Figures \ref{fig:5} (a-b) and (c-d) display the results by P-SMGP and BLDA, respectively. 
For Component 1, the target signal shows a major negative peak and a secondary positive peak around 300 and 500 ms post stimulus. The P-SMGP prior merges target and non-target transformed ERP responses during the first 160 ms and the last 200 ms. For Component 2, the target signal shows a major positive peak around 160 ms and a secondary negative peak around 400 ms. The P-SMGP prior merges target and non-target transformed ERP responses during the first 100 ms and the last 300 ms. The left and right columns of Figure \ref{fig:6} show the spatial pattern and spatial filter by xDAWN, respectively. For example, both spatial pattern and filters of the first component show that channels PO7, PO8, and Oz is a key contributor to the target neural activity. Since swLDA is a discriminative model rather than a generative one, it does not provide direct parameter estimates. In each figure, the solid red and blue lines represent the posterior mean estimates of $\bm{\beta}_{1}$ and $\bm{\beta}_{0}$, respectively, while the shaded areas illustrate their corresponding 95\% credible intervals.

Both P-SMGP and BLDA capture the overall temporal dynamics of $\bm{\beta}_{1}$ and $\bm{\beta}_{0}$, showing clear variations over time. However, the P-SMGP method exhibits two key advantages. First, its credible intervals are generally narrower than those of BLDA, indicating more precise parameter estimation with reduced uncertainty. Second, in time regions where the difference between $\bm{\beta}_{1}$ and $\bm{\beta}_{0}$ is small, P-SMGP ensures that their estimates nearly overlap. In contrast, BLDA not only produces wider credible intervals but also tends to separate the estimates of $\bm{\beta}_{1}$ and $\bm{\beta}_{0}$ even in regions where their differences should be minimal. 

\subsubsection{Prediction Performance}
\begin{figure}[htbp]
  \centering

  \begin{minipage}[b]{0.30\textwidth}
    \centering
    \includegraphics[width=\textwidth]{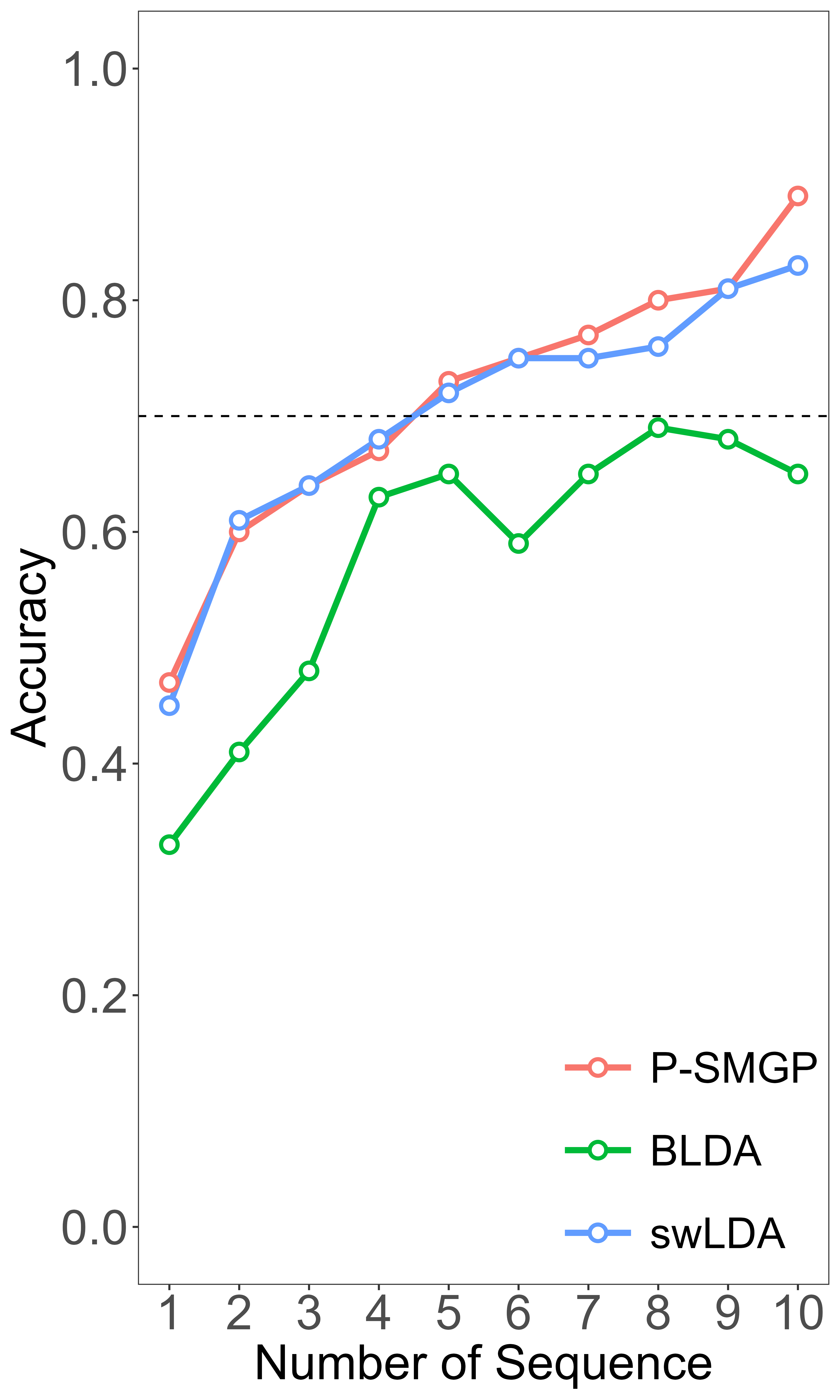}
    {\small (a)}
  \end{minipage}
  \hfill
  \begin{minipage}[b]{0.31\textwidth}
    \centering
    \includegraphics[width=\textwidth]{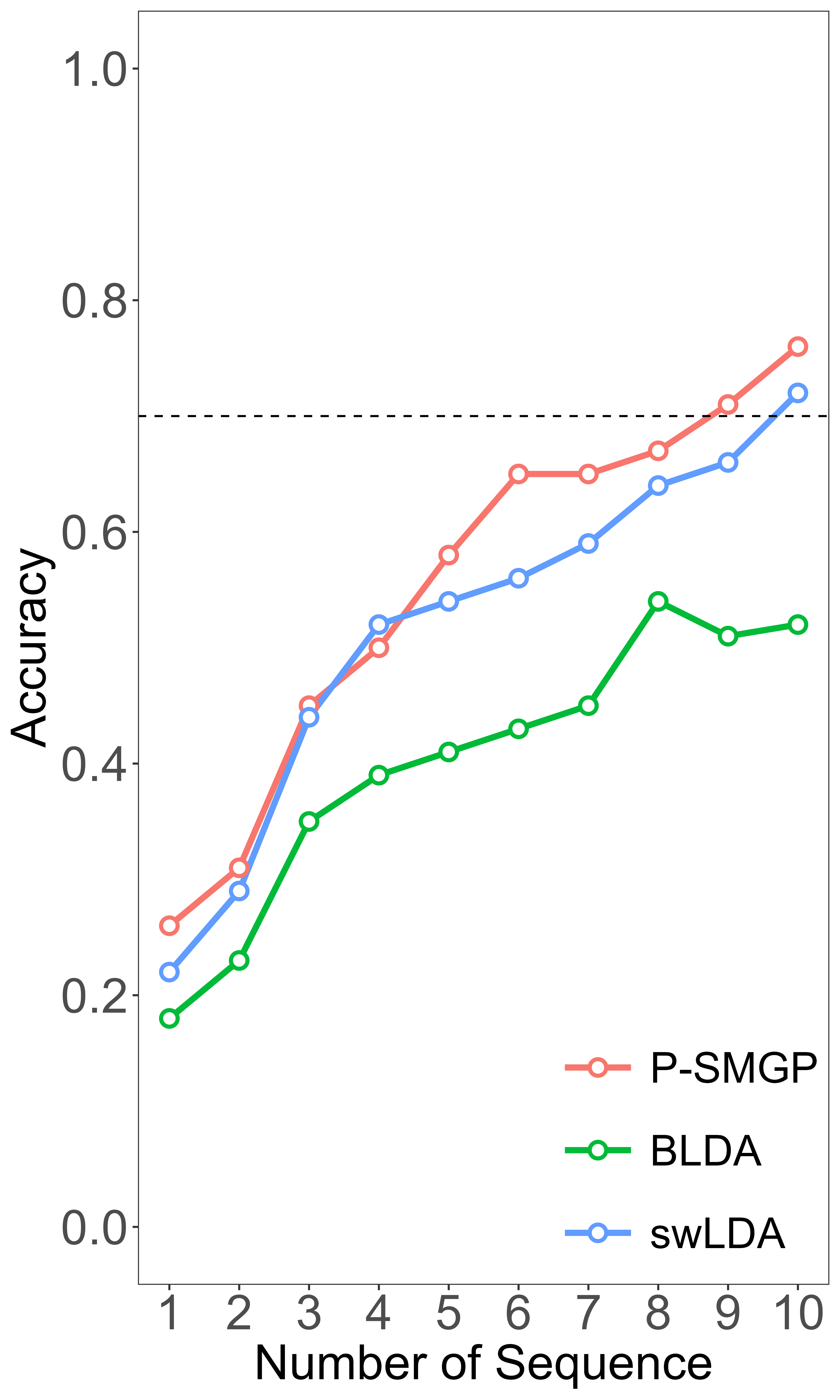}
    {\small (b)}
  \end{minipage}
  \hfill
  \begin{minipage}[b]{0.31\textwidth}
    \centering
    \includegraphics[width=\textwidth]{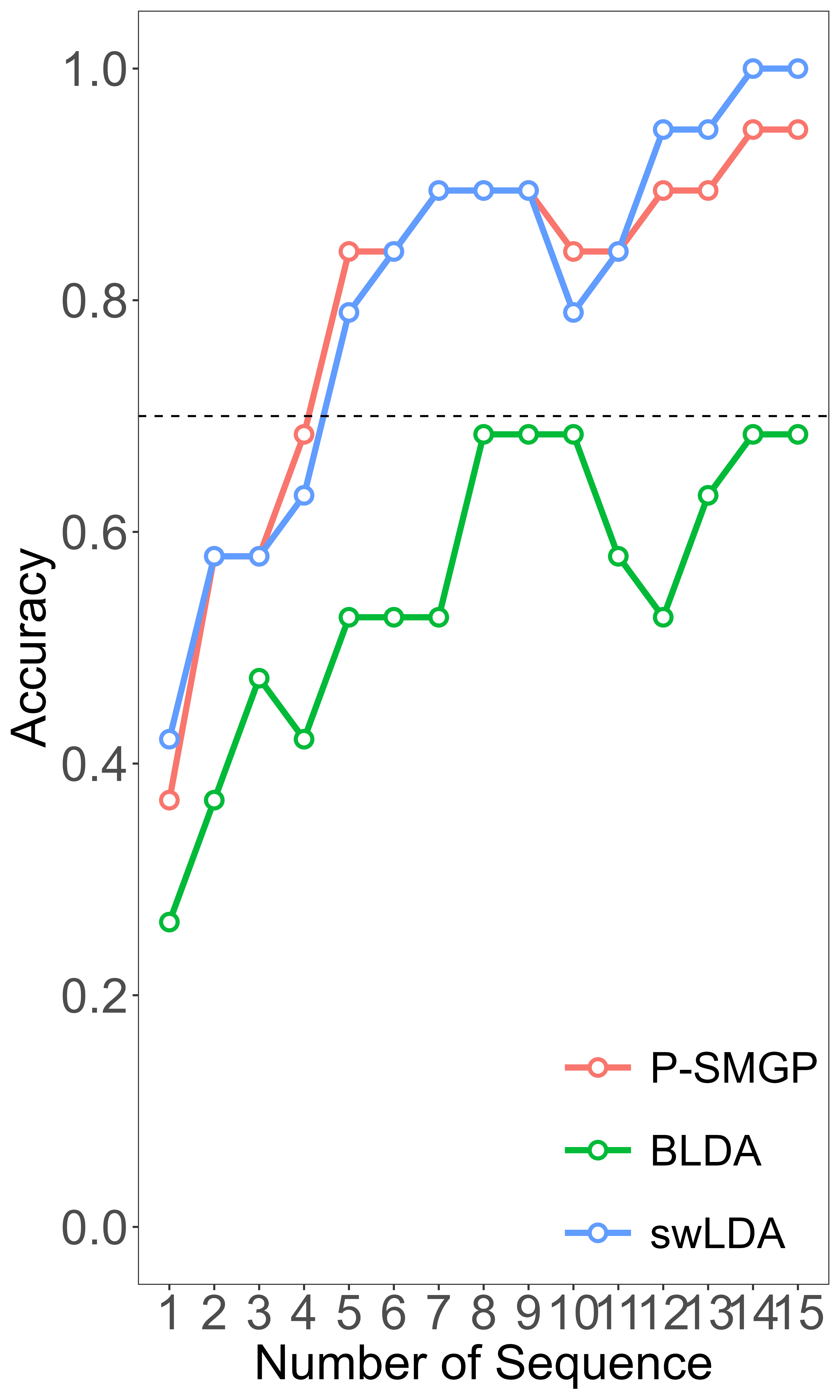}
    {\small (c)}
  \end{minipage}

  \caption{\small Prediction accuracy of P-SMGP, swLDA and BLDA for data from real participant K178 with different testing sequence sizes in (a) BCI (b) DYN (c) CMP scenarios. The dashed line indicates the critical 70\% accuracy threshold \cite{kubler_brain-computer_2005} for practical BCI usability.
  }
  \label{fig:7}
\end{figure}

To further evaluate the performance of different methods in real data analysis for K178, we compared their character-level prediction accuracy across different sequence sizes in three distinct scenarios: BCI, DYN, and CMP. Figure \ref{fig:7} summarizes the accuracy results for P-SMGP, swLDA, and BLDA.

Across all three scenarios, P-SMGP outperformed BLDA and performed slightly better than swLDA. In the BCI scenario, P-SMGP and swLDA performed similarly at smaller sequence sizes, but as more sequences accumulate (e.g., 6–10), P-SMGP maintained a slight advantage, achieving an 89\% accuracy at a sequence size of 10, while swLDA and BLDA showed prediction accuracy of 83\% and 65\%, respectively. A similar trend was observed in the DYN scenario (P-SMGP: 76\%, swLDA: 72\%, and BLDA: 52\%) and in the CMP scenario (P-SMGP: 78\%, swLDA: 76\%, and BLDA: 58\%) at sequence 10.

In addition, we reported the prediction accuracy of all 16 participants at their maximum sequence size across the three experimental scenarios in Table \ref{tab:2}. Overall, P-SMGP and swLDA exhibited similar classification accuracy, with both methods achieving high performance across most participants, while BLDA achieved the lowest prediction accuracy. In the BCI condition, P-SMGP and swLDA generally achieved comparable results, with individual accuracies varying between participants. Some participants, such as K151 and K183, achieved over 90\% accuracy with both P-SMGP and swLDA, while others, like K185 and K160, showed lower accuracy across all models. However, compared to BLDA, which exhibited substantial performance degradation in several cases, both P-SMGP and swLDA provided more reliable and stable classification. A similar trend was observed in the DYN scenario, where P-SMGP and swLDA again achieved closely matched results, with both consistently outperforming BLDA. For certain participants, such as K114 and K172, P-SMGP showed a slight advantage over swLDA, reaching near-perfect accuracy, while for others, such as K106 and K113, swLDA marginally outperformed P-SMGP. However, the difference between P-SMGP and swLDA remained relatively small, whereas BLDA showed a clear drop in accuracy across nearly all participants, reinforcing its weaker predictive capability. In the CMP scenario, P-SMGP and swLDA maintained their competitive performance, with both methods achieving comparable accuracy levels across participants.

\begin{table}[ht]
\centering
\small
\caption{Cumulative prediction accuracy of P-SMGP, swLDA and BLDA for data from real participants across different scenarios (BCI, DYN, and CMP). Participants where P-SMGP outperforms swLDA and BLDA methods in the BCI scenario are highlighted in bold.}
\label{tab:2}
\begin{tabularx}{\textwidth}{ 
  >{\centering\arraybackslash\hsize=1.8\hsize}X | 
  *{3}{>{\centering\arraybackslash\hsize=0.9\hsize}X} |
  *{3}{>{\centering\arraybackslash\hsize=0.9\hsize}X} |
  *{3}{>{\centering\arraybackslash\hsize=0.9\hsize}X} 
}
\hline
\multirow{2}{*}{ID} & \multicolumn{3}{c|}{BCI} & \multicolumn{3}{c|}{DYN} & \multicolumn{3}{c}{CMP} \\ \cline{2-10} 
                              & \mbox{P-SMGP}   & swLDA   & BLDA  & \mbox{P-SMGP}   & swLDA   & BLDA  & \mbox{P-SMGP}   & swLDA  & BLDA  \\ \hline
\textbf{K106}                          & 0.86   & 0.84    & 0.65  & 0.81   & 0.84    & 0.55  & 0.84   & 0.84   & 0.58  \\
K107                          & 0.51   & 0.56    & 0.41  & 0.58   & 0.67    & 0.21  & 0.44   & 0.54   & 0.17  \\
\textbf{K113}                          & 0.67   & 0.61    & 0.66  & 0.79   & 0.80    & 0.35  & 0.64   & 0.52   & 0.19  \\
K114                          & 0.88   & 0.93    & 0.83  & 0.96   & 0.99    & 0.86  & 0.84   & 0.94   & 0.84  \\
\textbf{K145}                          & 0.68   & 0.63    & 0.50  & 0.61   & 0.58    & 0.49  & 0.73   & 0.70   & 0.59  \\
K151                          & 0.96   & 0.97    & 0.97  & 0.94   & 0.96    & 0.41  & 0.92   & 0.92   & 0.49  \\
K154                          & 0.53   & 0.53    & 0.53  & 0.79   & 0.83    & 0.67  & 0.72   & 0.76   & 0.68  \\
\textbf{K159}                          & 0.84   & 0.83    & 0.82  & 0.76   & 0.77    & 0.72  & 0.75   & 0.76   & 0.76  \\
\textbf{K160}                          & 0.49   & 0.42    & 0.21  & 0.53   & 0.52    & 0.17  & 0.36   & 0.35   & 0.24  \\
K172                          & 0.67   & 0.70    & 0.68  & 0.80   & 0.80    & 0.22  & 0.72   & 0.77   & 0.07  \\
\textbf{K178}                          & 0.89   & 0.83    & 0.65  & 0.76   & 0.72    & 0.52  & 0.77   & 0.75   & 0.59  \\
K183                          & 0.93   & 0.97    & 0.93  & 0.88   & 0.89    & 0.87  & 0.90   & 0.95   & 0.88  \\
K184                          & 0.88   & 0.95    & 0.68  & 0.84   & 0.90    & 0.46  & 0.80   & 0.89   & 0.49  \\
K185                          & 0.34   & 0.48    & 0.34  & 0.35   & 0.36    & 0.34  & 0.41   & 0.42   & 0.40  \\
\textbf{K190}                          & 0.66   & 0.63    & 0.43  & 0.56   & 0.53    & 0.14  & 0.65   & 0.68   & 0.19  \\
K191                          & 0.56   & 0.59    & 0.58  & 0.50   & 0.56    & 0.50  & 0.48   & 0.44   & 0.44  \\ \hline
\end{tabularx}
\end{table}

\section{Discussion}

In this study, we introduced a novel approach to ERP analysis by modeling EEG signals on the stimulus level using Probit-link Split-and-Merge Gaussian Process (P-SMGP) for feature selection. This framework captured the temporal dynamics of target and non-target transformed ERP responses through merged latent functions ($\beta_{1}(t)$ and $\beta_{0}(t)$), enabling adaptive identification of discriminative time intervals. Unlike swLDA that uses non-interpretable feature selection, 
\textcolor{violet}{P-SMGP provides the interpretable temporal feature selection while maintaining the probabilistic uncertainty quantification.} 

Compared with the SMGP framework of \citet{ma_bayesian_2022}, our P-SMGP approach differs in several methodological aspects. First, while the original model was developed for sequence-level inference, we modeled the EEG signals on the stimulus level to facilitate the prediction and make direct inferences on the target and non-target ERP response functions. 
Second, instead of using the truncated normal formulation, we adopted a probit link to simplify the posterior inferences and
\textcolor{violet}{facilitate posterior computation.}
Third, rather than fitting the model to raw EEG signals, we based our analysis on transformed ERP data to improve the signal-to-noise ratio and make the response patterns more salient. The temporal features of the transformed EEG signals were still preserved.
Finally, for prediction, we applied a weighted-likelihood-based approach to update the character-level probability, as opposed to the direct summation of stimulus-specific classifier scores or the indirect manipulation on the sequence level.

The simulation results demonstrated that P-SMGP consistently outperformed swLDA and BLDA in both parameter estimation and classification accuracy. In parameter estimation, P-SMGP achieved superior precision characterized by narrower credible intervals, enabling clearer differentiation between target and non-target signal characteristics during feature selection. 
\textcolor{violet}{This improved precision may contribute to the observed prediction performance by providing clearer differentiation between target and non-target response patterns,}
as evidenced by the tight credible intervals observed in stable signal regions. The merging mechanism effectively selects temporal intervals where target-nontarget signal differences exceed probabilistic thresholds, corroborating our theoretical framework that selective interval merging mitigates overfitting in low-information regions.

In real-data analyses, P-SMGP replicated simulation advantages across three experimental scenarios (BCI, DYN, and CMP), achieving the critical 70\% accuracy threshold \cite{kubler_brain-computer_2005} for practical BCI usability at sequence size 10. Participant K178's results revealed 89\% BCI accuracy (vs. swLDA's 83\%), with P-SMGP maintaining less accuracy variance across sessions compared to swLDA's fluctuations. For cross-participant analyses, P-SMGP shows its robustness: median accuracy improvements of 14\% over BLDA (BCI: 82\% vs. 68\%) and comparable performance to swLDA. Meanwhile, our feature selection provides statistical evidence on the separation effects between target and non-target transformed ERP response functions.

Across participants, the two latent components learned by our model exhibit ERP-like temporal morphologies that are consistent with the classical responses used in P300-speller paradigms. Specifically, Component 1 typically reflects an early negative-going deflection (i.e., an N200-like pattern), whereas Component 2 reflects a later positive-going deflection (i.e., a P300-like pattern). This interpretation is supported by the observed waveform shapes and their characteristic latency ranges: Component 1 emphasizes earlier negative deviations, while Component 2 emphasizes later positive deviations around the target-related response window. Importantly, the model-selected (merged) time points tend to concentrate near the onset/offset or within these discriminative windows, indicating that the model is preferentially retaining time locations where N200/P300-related separability is strongest.

We also observe cross-participant similarity in the estimated component trajectories. For example, participants such as K114, K151, K183, K184, and K191 show a qualitatively similar pattern in which Component 1 remains relatively stable for a short period and then presents a pronounced negative deflection around the late response window, while Component 2 shows smaller fluctuations before stabilizing; their merged time points are likewise similar and often occur near the beginning or end of the signal. This suggests that participants may share partially transferable neuro-physiological structure in these ERP-related components. As future work, we plan to improve spelling efficiency by borrowing information from source participants, enabling a new participant profile to be learned with fewer calibration trials while preserving individual-specific variability.


\textcolor{violet}{To summarize, our findings highlighted the potential of the overall xDAWN--P-SMGP framework to provide competitive classification accuracy while offering interpretable temporal feature selection in ERP-based BCIs.} 
By addressing the limitations of existing classifiers and introducing a Bayesian generative alternative, this study contributed to the development of more robust and personalized BCI systems.

\section{Disclosure statement}
\label{disclosure-statement}

The authors declare no conflicts of interests.

Participants either signed consent forms approved by IRB or gave consent and had a caregiver sign on their behalf. More details can be found \if1\anon \cite{thompson_plug-and-play_2014}\fi \if0\anon XXX\fi.

\section{Data Availability Statement}
\label{data-availability-statement}

The participants in the study did not consent to release their data in public directly, even for the de-identified version. However, the de-identified data are available upon request for other researchers. According to the guideline from the 
\if1\anon
{University of Michigan medical school office of research}\fi
\if0\anon{XXX University}\fi, an outgoing data user agreement (DUA) needs to be created for external sharing of any individual-level clinical data, regardless of de-identified or not.

\phantomsection\label{supplementary-material}
\bigskip

\begin{center}

{\large\bf SUPPLEMENTARY MATERIAL}

\end{center}

\begin{description}
\item[Additional real data analysis] A PDF file to demonstrate additional results of the real data analysis.
\item[GitHub for codes] A link to the GitHub repository. \url{https://github.com/xiaozhaji-wu/P-SMGP_BCI_EEG}
\end{description}



\bibliography{bibliography}

\end{document}